\documentclass[10pt, twocolumn, nofootinbib,superscriptaddress]{revtex4-1}
\usepackage{amsmath,amssymb,amsfonts}
\usepackage{algorithmic}
\usepackage{graphicx}
\usepackage{textcomp}
\usepackage{xcolor}
\usepackage{ragged2e}
\usepackage{booktabs, makecell, tabularx}

\usepackage{gensymb}
\makeatletter
    \renewcommand\@make@capt@title[2]{%
     \@ifx@empty\float@link{\@firstofone}{\expandafter\href\expandafter{\float@link}}%
      {\textbf{#1}}\@caption@fignum@sep#2\quad}%
\makeatother
\makeatletter 
\renewcommand{\fnum@figure}{\textbf{Fig.~\thefigure}} 
\makeatother

\usepackage{xcolor}

\def\BibTeX{{\rm B\kern-.05em{\sc i\kern-.025em b}\kern-.08em
    T\kern-.1667em\lower.7ex\hbox{E}\kern-.125emX}}

\begin{document}

\author{Kaixuan Ye}
\thanks{These authors contributed equally to this work}
\affiliation{Nonlinear Nanophotonics Group, MESA+ Institute of Nanotechnology,\\
University of Twente, Enschede, Netherlands}
\author{Hanke Feng}
\thanks{These authors contributed equally to this work}
\affiliation{Department of Electrical Engineering $\&$ State Key Laboratory of Terahertz\\ and Millimeter Waves, City University of Hong Kong, Hong Kong, China}
\author{Yvan Klaver}
\affiliation{Nonlinear Nanophotonics Group, MESA+ Institute of Nanotechnology,\\
University of Twente, Enschede, Netherlands}
\author{Akshay Keloth}
\affiliation{Nonlinear Nanophotonics Group, MESA+ Institute of Nanotechnology,\\
University of Twente, Enschede, Netherlands}
\author{Akhileshwar Mishra}
\affiliation{Nonlinear Nanophotonics Group, MESA+ Institute of Nanotechnology,\\
University of Twente, Enschede, Netherlands}
\author{Cheng Wang}
\email{cwang257@cityu.edu.hk}
\affiliation{Department of Electrical Engineering $\&$ State Key Laboratory of Terahertz\\ and Millimeter Waves, City University of Hong Kong, Hong Kong, China}
\author{David Marpaung}
\email{david.marpaung@utwente.nl}
\affiliation{Nonlinear Nanophotonics Group, MESA+ Institute of Nanotechnology,\\
University of Twente, Enschede, Netherlands}

\date{\today}
\title{Surface acoustic wave stimulated Brillouin scattering in thin-film lithium niobate waveguides}

\begin{abstract}
We report the first-ever experimental observation of backward stimulated Brillouin scattering (SBS) in thin-film lithium niobate (TFLN) waveguides. The peak Brillouin gain coefficient of the z-cut LN waveguide with a crystal rotation angle of 20$\degree$ is as high as 84.9~m$^{-1}$W$^{-1}$, facilitated by surface acoustic waves (SAW) at 8.06~GHz.
\end{abstract}
\maketitle

Stimulated Brillouin scattering (SBS), a nonlinear optomechanical interaction involving light and GHz acoustic waves, has spurred groundbreaking applications \cite{Eggleton2019}, from sub-Hertz level laser \cite{Gundavarapu2018}, ultra-high selectivity filter \cite{Marpaung2015Low-powerSelectivity}, to optical gyroscope \cite{Lai2020}. Exploiting the SBS effect in chip-scale photonic devices for Brillouin-based applications requires a large Brillouin gain coefficient in a scalable photonic integrated platform. Remarkably, strong SBS effect with Brillouin gain coefficient over 100~m$^{-1}$W$^{-1}$ has been observed in chalcogenide \cite{Pant2011} and suspended-silicon \cite{Kittlaus2016} waveguides. While promising, these platforms are either volatile or mechanically unstable. On-chip SBS has also been explored in low-loss and scalable silicon nitride platforms \cite{Botter2022,Gundavarapu2018}. Nevertheless, Brillouin gain coefficients in these platforms are limited to below 1~m$^{-1}$W$^{-1}$.

Very recently, theoretical works predict a strong SBS effect in thin-film lithium niobate (TFLN) waveguides \cite{Rodrigues2023}. TFLN waveguides, recognized for their low loss, scalability, and versatility \cite{Boes2023}, have enabled unprecedented performances and functions in modulators \cite{Wang2018}, optical frequency combs \cite{Zhang2019}, and quantum optics \cite{McKenna2020}.  While the optomechanical effect has been investigated in TFLN waveguides, previous experiments rely on electric interdigital transducers to externally excite the acoustic waves \cite{Sarabalis2021,Shao2019}. The SBS effect in TFLN waveguides has never been observed.

Here, we report, to the best of our knowledge, the first-ever experimental observation of backward SBS signals from TFLN waveguides. Our investigation encompasses both z-cut and x-cut TFLN waveguides. Notably, we observed the SBS effect in air-cladded z-cut TFLN waveguides with various crystal rotation angles ($\theta$), confirming the crystal orientation dependence of the SBS strength. Furthermore, we experimentally demonstrated the enhancement of the SBS effect by surfaced acoustic waves by comparing x-cut TFLN waveguides with both silica and air cladding. The massive Brillouin gain coefficient (84.9~m$^{-1}$W$^{-1}$ in z-cut TFLN waveguides with $\theta$=20$\degree$) in the mature TFLN platform makes it an ideal candidate for integrated Brillouin-based applications.

\begin{figure*}[t!]
\centering
\includegraphics[width=\linewidth]{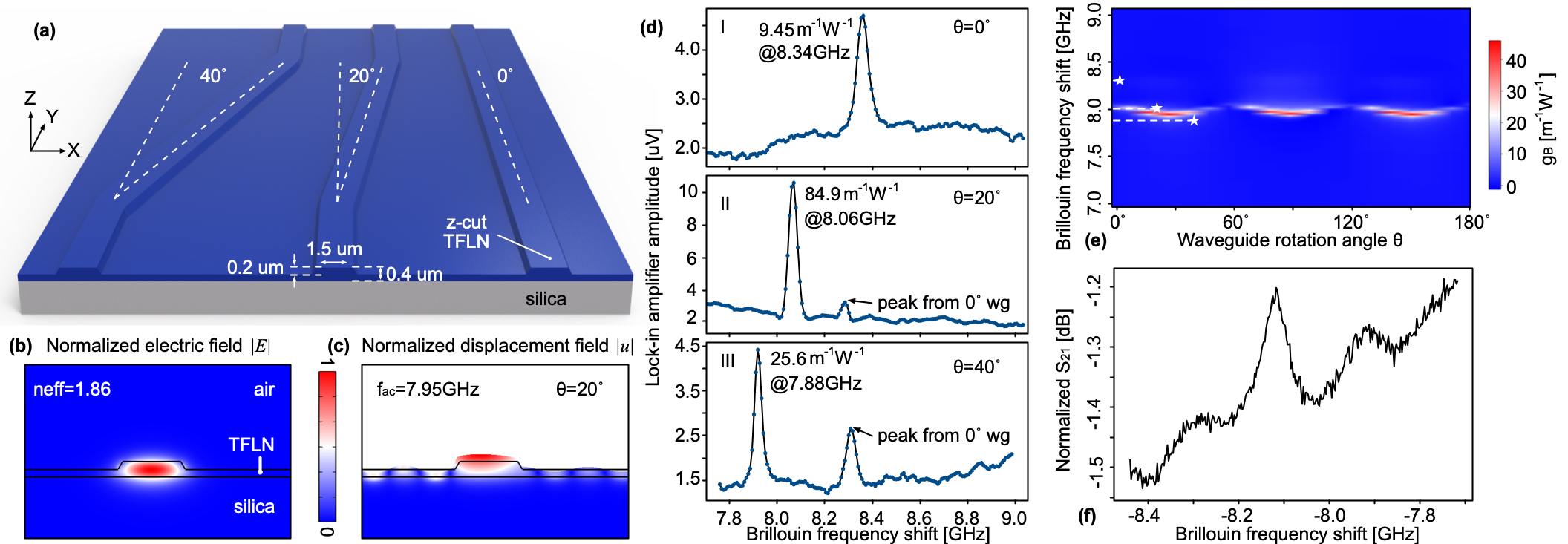}
\caption{(a) Artistic representation of the air-cladded z-cut TFLN waveguides with crystal orientation angle $\theta$ of 0$\degree$, 20$\degree$, and 40$\degree$. (b) Simulated electric field and (c) the corresponding acoustic response of the air-cladded z-cut TFLN waveguides at the Brillouin frequency shift ($\theta$ = 20$\degree$). (d) Lock-in amplifier measurement results of the Brillouin gain profile of TFLN waveguides with different rotation angles $\theta$. The Brillouin gain coefficients are calculated by comparing waveguide peaks with fiber peaks. (e) Simulated Brillouin gain density plot as a function of $\theta$. Stars indicate data points we measured. (f) SBS peak from the z-cut TFLN waveguides ($\theta$=~20$\degree$) from the VNA measurement. VNA: vector network analyzer.}
\label{fig1}
\end{figure*}

The air-cladded z-cut TFLN waveguides under test with $\theta$ = 0$\degree$, 20$\degree$, and 40$\degree$ are illustrated in Fig.\ref{fig1}~(a). We fabricate the waveguides using a commercial z-cut LN wafer (NANOLN) with a 400~nm-thick TFLN layer. The etch depth of the z-cut waveguide is 200~nm. All z-cut TFLN waveguides are 1-cm long. For waveguides with $\theta$ = 20$\degree$ and 40$\degree$, the angled section is 0.5~cm. All z-cut LN waveguides have a coupling loss of 5~dB per facet and a propagation loss of 0.8~dB/cm.

Fig.\ref{fig1}~(b) and (c) shows the simulated electric field and corresponding acoustic response of the z-cut TFLN waveguides with $\theta$~=~20$\degree$ in COMSOL. The waveguide supports surface acoustic waves at 7.95~GHz, resulting in a large overlap between the optical and acoustic fields for the SBS interaction.

We experimentally characterized the Brillouin gain profile of the z-cut TFLN waveguides with a pump-probe modulated lock-in amplifier setup \cite{Gyger2020,Botter2022,Botter2023}. Fig.\ref{fig1}~(d) shows the measured SBS peaks from the TFLN waveguides with different $\theta$.  We calculated the Brillouin gain coefficients by comparing the SBS peak from fibers in the setup (total length 5~m) and the waveguide peaks. Specifically, the z-cut TFLN waveguide at $\theta$=~0$\degree$ exhibits a Brillouin gain coefficient of 9.45 m$^{-1}$W$^{-1}$ at 8.34~GHz, while at $\theta$=~20$\degree$, the Brillouin gain coefficient reaches as high as 84.9~m$^{-1}$W$^{-1}$ at 8.06~GHz. For $\theta$=~40$\degree$, the Brillouin gain coefficient is reduced to 25.6~m$^{-1}$W$^{-1}$ at 7.88~GHz. The crystal orientation dependence of the SBS strength matches well with our simulation in Fig.\ref{fig1}~(e). Note that the simulated Brillouin gain coefficients are around 50\% lower than the measured values, which could be potentially due to the underestimated acoustic quality factor in our simulation model (Q$_{ac}$=1000).  Additionally, the SBS signal is further validated through vector network analyzer (VNA) measurement \cite{Botter2022,Botter2023}, as shown in Fig.\ref{fig1}~(f).

\begin{figure}[t!]
\centering
\includegraphics[width=\linewidth]{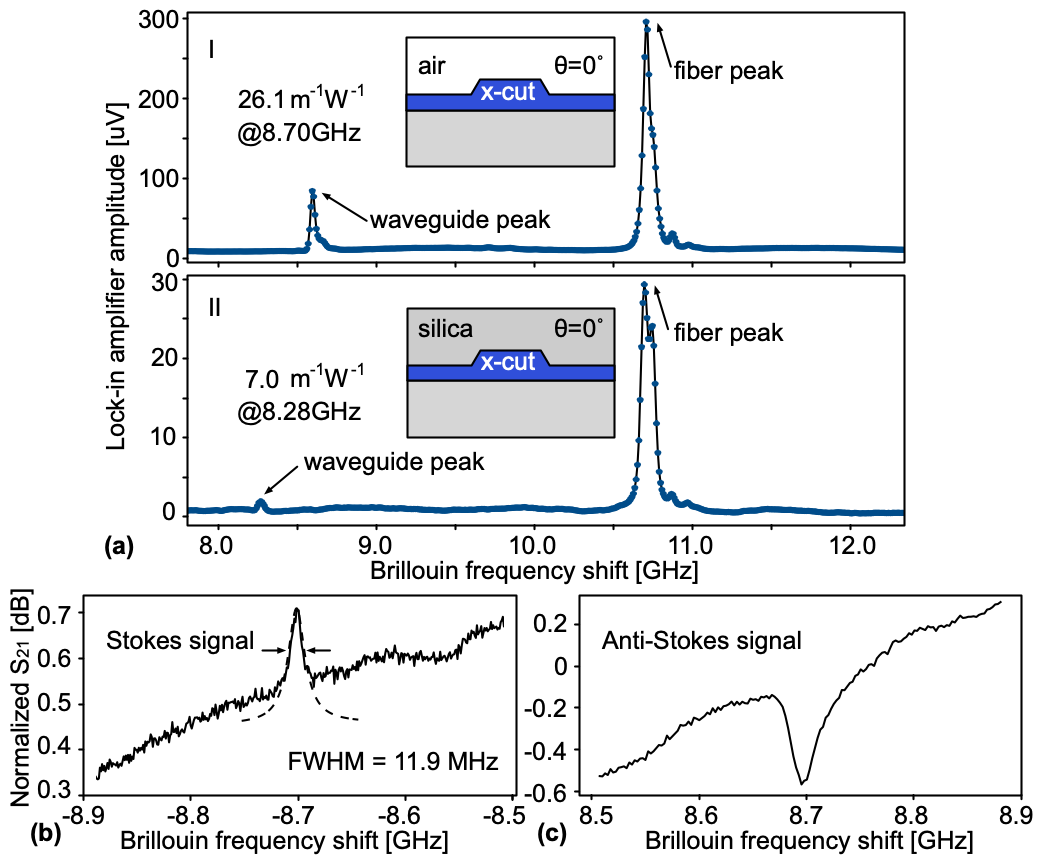}
\caption{(a)Lock-in amplifier measurement results of the x-cut TFLN waveguides with and without silica cladding, displayed together with SBS peaks from 5~m of fibers in the setup. Measured (b) Stokes and (c) anti-Stokes signal from the x-cut TFLN waveguides without silica cladding.}
\label{fig2}
\end{figure}

We also characterized Brillouin gain profiles of x-cut TFLN waveguides with an LN layer thickness of 500~nm and an etching depth of 250~nm. As shown in Fig.\ref{fig2}~(a), the measured Brillouin gain coefficient of the x-cut TFLN waveguide (2.8~cm long) with $\theta$ = 0$\degree$ is 26.1~m$^{-1}$W$^{-1}$ at 8.70~GHz. For further verification, we have also characterized the sample with the VNA measurement, as shown in Fig.\ref{fig2}~(b) and (c). Notably, the measured Stokes signal revealed a narrow linewidth of 11.9~MHz. To verify the role of surface acoustic waves in facilitating the SBS process, we also characterized samples with silica cladding on top of the x-cut TFLN waveguides (1~cm long). The Brillouin gain coefficients of the silica-cladded sample were reduced by nearly fourfold, primarily due to the diminished confinement of the acoustic wave compared to the air-cladded counterparts.

In conclusion, we have observed the first-ever SBS signal from TFLN waveguides in both z- and x-cut. The substantial Brillouin gain coefficients highlight the immense potential for leveraging SBS in the mature TFLN platform. This opens avenues for intersecting integrated Brillouin lasers, as well as Brillouin-based microwave photonics signal processing with existing functionalities such as modulators and optical frequency combs in the TFLN platform with unparalleled performances.

\bigskip
\section*{Author Contribution}
D.M. and K.Y. developed the concept and proposed the physical system. K.Y. and H.F. designed the photonic circuits. H.F. fabricated the photonic circuits. K.Y. and Y.K. developed and performed numerical simulations. K.Y. and Y.K. performed the experiments with input from A.K., and A.M.. K.Y. and D.M. wrote the manuscript with input from everyone. D.M. and C.W. led and supervised the entire project.
\label{sec:four}

\begin{acknowledgments}
The authors acknowledge funding from the European Research Council Consolidator Grant (101043229 TRIFFIC), Nederlandse Organisatie voor Wetenschappelijk Onderzoek (NWO) Start Up (740.018.021), the Research Grants
Council, University Grants Committee (N\_CityU113\/20, CityU11204022), and Croucher Foundation (9509005).
\end{acknowledgments}

\section*{Disclosures}
The authors declare no conflicts of interest. During the preparation of this manuscript, we notice a similar work by Caique C. Rodrigues et al. \cite{rodrigues2023onchip} also reports the SBS process in the TFLN platform.

\section*{Data Availability}
The data of this study are available
from the corresponding authors upon reasonable request.

\bibliographystyle{IEEEtran}
\bibliography{library}

\begin{thebibliography}{10}
\providecommand{\url}[1]{#1}
\csname url@samestyle\endcsname
\providecommand{\newblock}{\relax}
\providecommand{\bibinfo}[2]{#2}
\providecommand{\BIBentrySTDinterwordspacing}{\spaceskip=0pt\relax}
\providecommand{\BIBentryALTinterwordstretchfactor}{4}
\providecommand{\BIBentryALTinterwordspacing}{\spaceskip=\fontdimen2\font plus
\BIBentryALTinterwordstretchfactor\fontdimen3\font minus \fontdimen4\font\relax}
\providecommand{\BIBforeignlanguage}[2]{{%
\expandafter\ifx\csname l@#1\endcsname\relax
\typeout{** WARNING: IEEEtran.bst: No hyphenation pattern has been}%
\typeout{** loaded for the language `#1'. Using the pattern for}%
\typeout{** the default language instead.}%
\else
\language=\csname l@#1\endcsname
\fi
#2}}
\providecommand{\BIBdecl}{\relax}
\BIBdecl

\bibitem{Eggleton2019}
B.~J. Eggleton, C.~G. Poulton, P.~T. Rakich, M.~J. Steel, and G.~Bahl, ``Brillouin integrated photonics,'' \emph{Nature Photonics}, vol.~13, pp. 664--677, 2019.

\bibitem{Gundavarapu2018}
S.~Gundavarapu, G.~M. Brodnik, M.~Puckett, T.~Huffman, D.~Bose, R.~Behunin, J.~Wu, T.~Qiu, C.~Pinho, N.~Chauhan, J.~Nohava, P.~T. Rakich, K.~D. Nelson, M.~Salit, and D.~J. Blumenthal, ``Sub-hertz fundamental linewidth photonic integrated brillouin laser,'' \emph{Nature Photonics}, vol.~13, pp. 60--67, 12 2018.

\bibitem{Marpaung2015Low-powerSelectivity}
D.~Marpaung, B.~Morrison, M.~Pagani, R.~Pant, D.-Y. Choi, B.~Luther-Davies, S.~J. Madden, and B.~J. Eggleton, ``Low-power, chip-based stimulated brillouin scattering microwave photonic filter with ultrahigh selectivity,'' \emph{Optica}, vol.~2, p.~76, 2015.

\bibitem{Lai2020}
Y.~H. Lai, M.~G. Suh, Y.~K. Lu, B.~Shen, Q.~F. Yang, H.~Wang, J.~Li, S.~H. Lee, K.~Y. Yang, and K.~Vahala, ``Earth rotation measured by a chip-scale ring laser gyroscope,'' \emph{Nature Photonics}, vol.~14, pp. 345--349, 2 2020.

\bibitem{Pant2011}
R.~Pant, C.~G. Poulton, D.-Y. Choi, H.~Mcfarlane, S.~Hile, E.~Li, L.~Thevenaz, B.~Luther-Davies, S.~J. Madden, and B.~J. Eggleton, ``On-chip stimulated brillouin scattering,'' \emph{Optics Express}, vol.~19, pp. 8285--8290, 4 2011.

\bibitem{Kittlaus2016}
E.~A. Kittlaus, H.~Shin, and P.~T. Rakich, ``Large brillouin amplification in silicon,'' \emph{Nature Photonics}, vol.~10, pp. 463--467, 6 2016.

\bibitem{Botter2022}
R.~Botter, K.~Ye, Y.~Klaver, R.~Suryadharma, O.~Daulay, G.~Liu, J.~van~den Hoogen, L.~Kanger, P.~van~der Slot, E.~Klein, M.~Hoekman, C.~Roeloffzen, Y.~Liu, and D.~Marpaung, ``Guided-acoustic stimulated brillouin scattering in silicon nitride photonic circuits,'' \emph{Science Advances}, vol.~8, 2022.

\bibitem{Rodrigues2023}
C.~C. Rodrigues, R.~O. Zurita, T.~P.~M. Alegre, and G.~S. Wiederhecker, ``Stimulated brillouin scattering by surface acoustic waves in lithium niobate waveguides,'' \emph{Journal of the Optical Society of America B}, vol.~40, pp. 56--63, 2023.

\bibitem{Boes2023}
A.~Boes, L.~Chang, C.~Langrock, M.~Yu, M.~Zhang, Q.~Lin, M.~Lončar, M.~Fejer, J.~Bowers, and A.~Mitchell, ``Lithium niobate photonics: Unlocking the electromagnetic spectrum,'' \emph{Science}, vol. 379, 1 2023.

\bibitem{Wang2018}
C.~Wang, M.~Zhang, X.~Chen, M.~Bertrand, A.~Shams-Ansari, S.~Chandrasekhar, P.~Winzer, and M.~Lončar, ``Integrated lithium niobate electro-optic modulators operating at cmos-compatible voltages,'' \emph{Nature}, vol. 562, pp. 101--104, 9 2018.

\bibitem{Zhang2019}
M.~Zhang, B.~Buscaino, C.~Wang, A.~Shams-Ansari, C.~Reimer, R.~Zhu, J.~M. Kahn, and M.~Lončar, ``Broadband electro-optic frequency comb generation in a lithium niobate microring resonator,'' \emph{Nature}, vol. 568, pp. 373--377, 3 2019.

\bibitem{McKenna2020}
T.~P. McKenna, J.~D. Witmer, R.~N. Patel, W.~Jiang, R.~V. Laer, P.~Arrangoiz-Arriola, E.~A. Wollack, J.~F. Herrmann, and A.~H. Safavi-Naeini, ``Cryogenic microwave-to-optical conversion using a triply resonant lithium-niobate-on-sapphire transducer,'' \emph{Optica}, vol.~7, pp. 1737--1745, 12 2020.

\bibitem{Sarabalis2021}
C.~J. Sarabalis, R.~V. Laer, R.~N. Patel, Y.~D. Dahmani, W.~Jiang, F.~M. Mayor, and A.~H. Safavi-Naeini, ``Acousto-optic modulation of a wavelength-scale waveguide,'' \emph{Optica}, vol.~8, pp. 477--483, 4 2021.

\bibitem{Shao2019}
L.~Shao, M.~Yu, S.~Maity, N.~Sinclair, L.~Zheng, C.~Chia, A.~Shams-Ansari, C.~Wang, M.~Zhang, K.~Lai, and M.~Lončar, ``Microwave-to-optical conversion using lithium niobate thin-film acoustic resonators,'' \emph{Optica}, vol.~6, pp. 1498--1505, 12 2019.

\bibitem{Gyger2020}
F.~Gyger, J.~Liu, F.~Yang, J.~He, A.~S. Raja, R.~N. Wang, S.~A. Bhave, T.~J. Kippenberg, and L.~Thévenaz, ``Observation of stimulated brillouin scattering in silicon nitride integrated waveguides,'' \emph{Physical Review Letters}, vol. 124, 2020.

\bibitem{Botter2023}
R.~A. Botter, Y.~Klaver, R.~te~Morsche, B.~L.~S. Frare, B.~Hashemi, K.~Ye, A.~Mishra, R.~B.~G. Braamhaar, J.~D.~B. Bradley, and D.~Marpaung, ``Stimulated brillouin scattering in tellurite-covered silicon nitride waveguides,'' \emph{arXiv:2307.12814}, 7 2023.

\bibitem{rodrigues2023onchip}
C.~C. Rodrigues, N.~J. Schilder, R.~O. Zurita, L.~S. Magalhães, A.~Shams-Ansari, T.~P.~M. Alegre, M.~Lončar, and G.~S. Wiederhecker, ``On-chip backward stimulated brillouin scattering in lithium niobate waveguides,'' \emph{arXiv:2311.18135}, 2023.

\end{thebibliography}
\end{document}